\documentclass[aps,prl,twocolumn,floatfix,showpacs,preprintnumbers,superscriptaddress]{revtex4}

\usepackage{graphics}
\usepackage{graphicx}
\usepackage{epsfig}

\newcommand{\beq}{\begin{equation}}
\newcommand{\eeq}{\end{equation}}
\newcommand{\bea}{\begin{eqnarray}}
\newcommand{\eea}{\end{eqnarray}}
\usepackage{dcolumn}% Align table columns on decimal point
\usepackage{bm}% bold math%

\begin{document}

\title{Interactions of Spin Waves with a Magnetic Vortex}

\author{J.~P. Park}
\author{P.~A. Crowell}
  \email[]{crowell@physics.umn.edu}

\affiliation{School of Physics and Astronomy, University of Minnesota, 116 Church St. SE\\
Minneapolis, MN 55455}

\begin{abstract}
We have investigated azimuthal spin-wave modes in magnetic vortex structures using time-resolved Kerr microscopy.
Spatially resolved phase and amplitude spectra of ferromagnetic disks with diameters from 5 $\mu$m down to 500 nm
reveal that the lowest order azimuthal spin wave mode splits into a doublet as the disk size decreases.   We
demonstrate that the splitting is due to the coupling between spin waves and the gyrotropic motion of the vortex core.
\end{abstract}
\pacs{75.30.Ds, 75.75.+a, 75.40.Gb}
\maketitle

The magnetic ground state of a soft ferromagnetic particle is determined by a balance between magnetostatic and
exchange energies.  In the case of disks with negligible magnetocrystalline anisotropy, the stable configuration for
radii of several hundred nanometers and thicknesses of the order of 20 - 50 nm is a vortex with a core diameter on the
order of the exchange length\cite{Cowburn:PRL1999,Shinjo:Science2000}.  The magnetization outside of the core
circulates around the central axis, reducing the total magnetostatic energy.  This comes at the cost of a large
exchange energy near the vortex core, inside of which the magnetization rotates out of the plane of the disk. Because
the vortex state represents the simplest type of domain configuration that can be created in a uniformly magnetized
particle, its excitation spectra is of fundamental interest.  Recent experimental work has identified two classes of
excitations in magnetic vortices.  First, low-order magnetostatic spin waves can be observed by either time-resolved
Kerr microscopy\cite{Park:PRB2003,Buess:PRL2004,Buess:PRL2005,Zhu:PRB2005} or Brillouin
scattering\cite{Novosad:PRB2002,Giovannini:PRB2004}.  A second type of excitation is associated with the translational
degree of freedom of the vortex core itself\cite{Park:PRB2003,Choe:Science2004,Novosad:PRB2005}. This has a much lower characteristic frequency than the other
spin-wave modes, and it is for this reason that the two types of excitations are often treated as distinct.   For
example, the radial spin wave modes of a particle can be calculated accurately even if the position of the vortex core
is regarded as fixed\cite{Buess:PRL2004,Buess:PRL2005}.

In this Letter we report on a study of the dynamics of the lowest order azimuthal spin wave modes in ferromagnetic disks with diameters
between 500~nm and 5~$\mu$m.  In larger
diameter disks, these are degenerate as expected for a system with cylindrical symmetry.  We demonstrate that this
degeneracy is lifted in smaller disks (with diameters less than 2~$\mu$m and thickness to diameter ratio $> 0.005$) due
to the motion of the vortex core.  The relative phases of the two modes are determined by the polarity of the vortex
core, and the magnitude of the splitting is of the same order as the vortex gyrotropic frequency.  These results
demonstrate a significant coupling between vortex dynamics and magnetostatic spin waves.

Two sets of permalloy disks were fabricated on Si substrates using electron-beam lithography and electron-beam
evaporation of permalloy (Ni$_{0.81}$Fe$_{0.19}$). The first set of disks were used primarily for investigations of the
vortex gyrotropic mode and were 50 nm thick with diameters $D$ from 500 nm to 2 $\mu$m. The disks used for the detailed
investigation of azimuthal spin-wave modes covered the range of diameters from 700 nm to 5 $\mu$m and were 20 nm thick.
Each disk exhibited a single-vortex ground state as determined by magnetic force microscopy.
\begin{figure}
    \epsfxsize=8 cm
    \centerline{\epsfbox{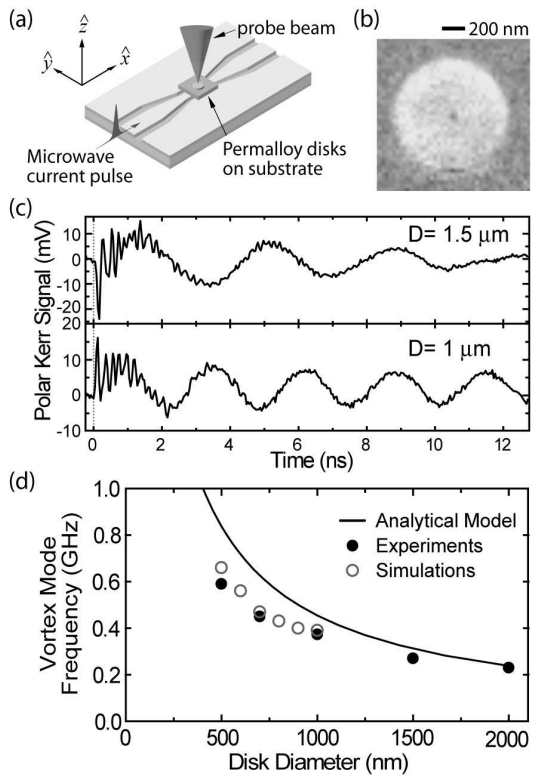}}
    \caption{(a) Schematic of the experiment showing the sample on a coplanar waveguide. The current pulse generates a magnetic field pulse along $\hat{y}$. (b) A magnetic force microscopy image of a $D$= 1$\mu$m disk showing a magnetic
vortex core at the center of the disk. (c) The polar Kerr signal as a function of pump-probe time delay measured for
disks with diameters of 1.5 and 1 $\mu$m. (d) The vortex mode eigenfrequency as a function of the disk diameter. The
analytical model (from Ref.~\onlinecite {Guslienko:JAP2002}) and simulation results are discussed in the text.}
    \label{fig:fig1}
\end{figure}

Measurements were conducted using time-resolved Kerr microscopy (TRKM)\cite{Hiebert:PRL1997,Park:PRL2002}. Each
substrate was thinned to a thickness of approximately 25 $\mu$m and mounted over the 30 $\mu$m wide center conductor of
a coplanar waveguide, as shown in Fig.\ref{fig:fig1}(a). The waveguide was placed on the scanning stage of an
oil-immersion microscope. The TRKM technique probes the response of the magnetization to a fast magnetic field pulse
$h_{p}$, which is along $\hat{y}$ in Fig.\ref{fig:fig1}(a). The magnitude of $h_p$ is 5 - 10 Oe, and the temporal width
of the pulse is 120 - 150 psec. The polar Kerr rotation $\theta(x,y;t)$ of an optical probe pulse is measured as a
function of the time delay $t$ between $h_p$ and the subsequent arrival of the probe pulse. A lock-in technique is used
to measure the change $\Delta\theta(x,y,t)$, due to $h_p$, averaged over many pulses. The z-component $m_z(x,y;t)$ of
the dynamic magnetization is proportional to $\Delta\theta(x,y;t)$.

In the remanent state, the vortex core is located at the center of the disk, as shown in the magnetic force micrograph
of  Fig.~\ref{fig:fig1}(b).  When the pulse $h_{p}$ is applied, the core is displaced and then slowly gyrates about the
center of the particle as the system relaxes towards equilibrium\cite{Park:PRB2003,Choe:Science2004}. This mode, which has also been observed in a resonance experiment\cite{Novosad:PRB2005},  appears as the long-lived low frequency ($< 1$~GHz) oscillation in Fig. \ref{fig:fig1}(c), which
shows the polar Kerr signal measured at a position 250 nm away from the center of two disks with diameters of 1.5 and 1
$\mu$m. The evolution of the vortex mode eigenfrequency as a function of the disk diameter is shown in Fig.
\ref{fig:fig1}(d).  As discussed in Ref.~\onlinecite {Guslienko:JAP2002}, the vortex core oscillates in an effective
potential that is due to the magnetostatic energy of the displaced vortex.   The solid curve shown in Fig.~\ref{fig:fig1}(d) follows from a model in which the vortex deforms rapidly (i.e. on a time scale fast relative to the core motion) in order to
cancel out the edge charges\cite{Guslienko:JAP2002}.  The  experimental results
are also in good agreement with a semi-analytical solution of the Landau-Lifshitz equation and micromagnetic
simulations\cite{Ivanov:JAP2004,Guslienko:PRB2005}.  The open circles in Fig.~\ref{fig:fig1}(d) show the
eigenfrequencies found in micromagnetic simulations of the full Landau-Lifshitz-Gilbert equation with a saturation magnetization $M_s = 770$~emu/cm$^3$, damping
parameter $\alpha=0.008$ and a cell size of 5 nm\cite{OOM:11b}.
\begin{figure}
    \epsfxsize=7.5 cm
    \centerline{\epsfbox{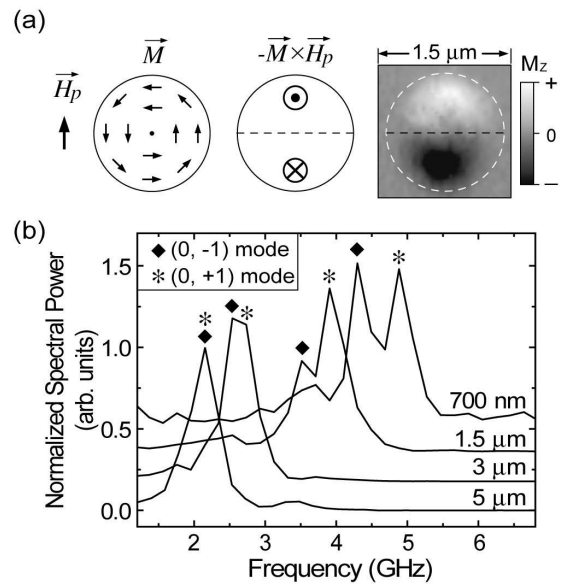}}
    \caption{(a) Schematic of the vortex magnetization (left), the torque exerted by the in-plane pulse (middle), and polar Kerr image immediately after the pulse for a 1 $\mu$m diameter disk (right).  (b) The spectral power measured for four disk diameters with thicknesses of 20 nm. The curves are offset vertically for clarity.}
    \label{fig:fig2}
\end{figure}

An additional feature of the data in Fig.~\ref{fig:fig1}(c) is the higher frequency response at short timescales. As illustrated
in Fig.~\ref{fig:fig2}(a), the torque ${\bf M}\times{\bf H}_p$ has opposite signs in the two halves of a vortex. As a
result, the spins in the upper half of the disk are rotated out of the plane, while those in the lower half of the disk
are rotated into the plane.  A spatially resolved polar Kerr image of a 1~$\mu$m diameter disk just after the pulse,
demonstrating the expected contrast, is shown in the right panel of Fig.~\ref{fig:fig2}(a).  The signal at short
time-scales in Fig.~\ref{fig:fig1}(c) is the subsequent precession and decay of the non-equilibrium magnetization.  The
spectrum of these higher frequency spin wave modes can be obtained by Fourier transformation of the time-domain data. Spectra
obtained for several different disk diameters are shown in Fig.~\ref{fig:fig2}(b).

Given the cylindrical symmetry of the vortex state, the spin wave modes can be indexed by the number of nodes $n$ and
$m$ along the radial $\hat{r}$ and azimuthal $\hat{\phi}$ directions: $(n,m)\equiv {\bf m}_{n,m}(r;t)
e^{i(m\phi+\omega_{n,m}t)}$, where $f_{(n,m)}=\omega_{n,m}/2\pi$ is the eigenfrequency.  This enumeration ignores the
node that always exists at the vortex core, inside of which the in-plane
components of the magnetization vanish.  The (0,0) mode is a purely radial mode and is excited most efficiently by an
out-of-plane magnetic field pulse as observed in Ref.~\onlinecite{Buess:PRL2004}.  The (0,$\pm$1) modes are the two
lowest order azimuthal modes and are excited by an in-plane pulsed
field\cite{Buess:PRL2004,Zaspel:PRB2005,Zhu:PRB2005}. Because of the cylindrical symmetry of  the vortex ground state,
the (0,$\pm$1) modes are expected to be degenerate if the core is fixed at the center of the disk.  As expected, only a
single peak is observed in the spectra of larger disks, as shown for the case of a  5 $\mu$m diameter disk in Fig.
\ref{fig:fig2}(b).
\begin{figure}
    \epsfxsize=8.2 cm
    \centerline{\epsfbox{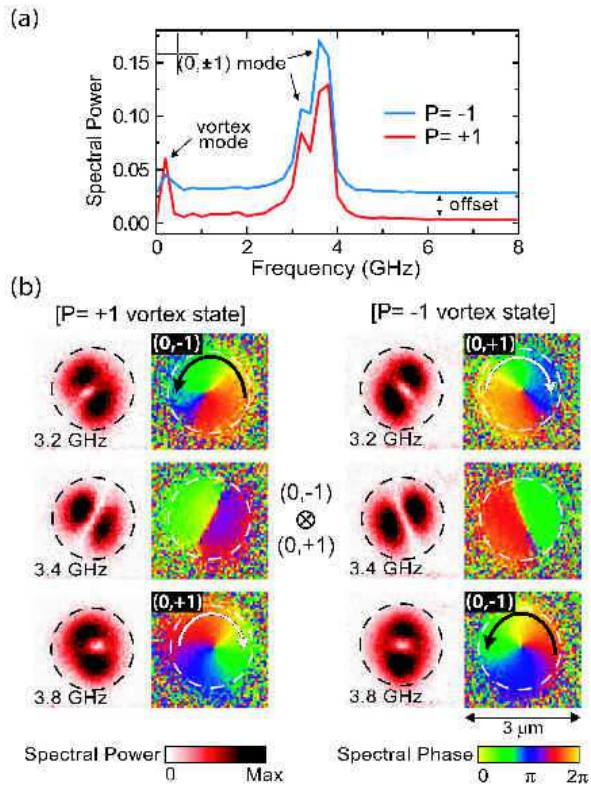}}
    \caption{(color on-line) (a) The spectral power integrated over space for two different vortex-core polarities: blue (top) for the $P=-1$ vortex state and red (bottom) for the $P=+1$ state of a 1.5~$\mu$m diameter disk.  The chiralities of the two vortex states are the same.  The two curves are offset for clarity. (b) The spectral power and phase images of the (0,$\pm$1) modes at 3.2 GHz (top) and 3.8 GHz (bottom).  The panels on the left show results for the $P=+1$ vortex state (power and phase, respectively) and the right panels the $P=-1$ state. The arrows show the direction in which the phase changes from 0 to $2\pi$.  The response at 3.4 GHz (middle) is the superposition of the (0,-1) and (0,+1) modes.}
    \label{fig:fig3}
\end{figure}

Although a single (0,$\pm$1) mode is observed in the 5 $\mu$m diameter disk, the peaks in Fig.~\ref{fig:fig2}(b) begin
to split as the disk diameter decreases.  This doublet has also been observed in measurements on a 700 nm diameter disk
by Zhu {\it et al.}\cite {Zhu:PRB2005} The remainder of this paper is dedicated to understanding these doublets.
Clearly the lifting of the degeneracy requires the breaking of cylindrical symmetry, and we argue that the reduced
symmetry originates from the gyrotropic motion of the vortex core, an argument supported by more rigorous
calculations\cite{Ivanov:PRL2005,Zaspel:PRB2005} and micromagnetic simulations\cite{Zhu:PRB2005}. To address this we
will label the vortices by a polarity $P=\pm 1$, where $P$ corresponds to the vortex core pointing out of ($P = +1$) or
into ($P=-1$) the plane of the disk, and a chirality $C=\pm 1$ indicating that the in-plane component of the
magnetization rotates counterclockwise ($C = +1$) or clockwise ($C = -1$).

As noted above, the low-frequency vortex mode is gyrotropic and therefore has a particular sense of rotation. Ignoring
damping, the total force acting on the vortex is\cite{Thiele:PRL1973}
\begin{equation}\label{eq:Thiele}
\mathbf{F} = \mathbf{G}\times \frac{d\mathbf{R}}{dt} - \frac{\partial W(\mathbf{R})}{\partial{\mathbf R}},
\end{equation}
where ${\bf R}$ is the position of the core in the plane, ${\bf G}$ is the gyrovector with a direction (into or out of
the plane) corresponding to the sign of the polarization $P$, and $W$ is the total energy of the vortex.  Since $W$
does not depend on the chirality, the gyrotropic motion described by Eq.~\ref{eq:Thiele} depends only the vortex
polarity $P$.  In our experiment, the
core polarity can be controlled by relaxing the disk from saturation in the presence of a small ($\sim 80$~Oe)
out-of-plane field.  The chirality cannot be controlled but is easily measured by imaging the displacement of the
vortex in a small in-plane field.  (The direction of displacement is perpendicular to the applied field and depends on
the chirality.) In this way, we can select two vortices with identical chiralities but opposite polarities.  Spectra
for these two vortex states in a 1.5 $\mu$m disk are shown in Fig.~\ref{fig:fig3}(a).   Each spectrum shows three peaks
at identical frequencies of 0.2, 3.2, and 3.8 GHz.  The upper two modes are the (0,$\pm$1) modes, for which a splitting
can be resolved.

Fig.~\ref{fig:fig3}(b) shows the spectral power and phase images at three different frequencies for the two vortex
polarities.  These images are constructed from the Fourier transform of the time-domain data\cite
{Park:PRB2003,Buess:PRL2004}.  As expected for azimuthal modes with one node, the phase wraps through 2$\pi$ as the
disk is traversed around the circumference.  The spectral power images at the two peak frequencies (3.2 and 3.8 GHz),
shown in the top and bottom panels of Fig.~\ref{fig:fig3}(b),  are essentially the same for the two polarities.  The
phase images, however, invert when the polarity of the vortex core is reversed.  In other words, the direction in which
the phase winds from $0\rightarrow 2\pi$ changes from counterclockwise to clockwise.  This corresponds to reversing
the sign of the azimuthal index $m$, and so the (0,+1) and (0,-1) modes exchange identities when the polarity of the
vortex core is reversed.

It is also possible to examine the spectral power and phase in the frequency
regime corresponding to a superposition of the two modes.  This is shown in the middle panel of Fig.~\ref{fig:fig3}(b).
An equal superposition of the two $(0,\pm 1)$ modes would be a standing wave, and indeed a line of nodes can be seen in
the spectral power image at 3.4 GHz.  Since the wave vectors of the (0,$\pm 1$) modes are either parallel or anti-parallel to the magnetization, they have
backward volume character\cite{BWVMS}, and both eigenfrequencies are therefore expected to lie {\it below} that of the
radially symmetric (0,0) mode\cite{Buess:PRB2005,Zhu:PRB2005}, although the (0,0) mode cannot be excited in the geometry of this experiment.

\begin{figure}
    \epsfxsize=8 cm
    \centerline{\epsfbox{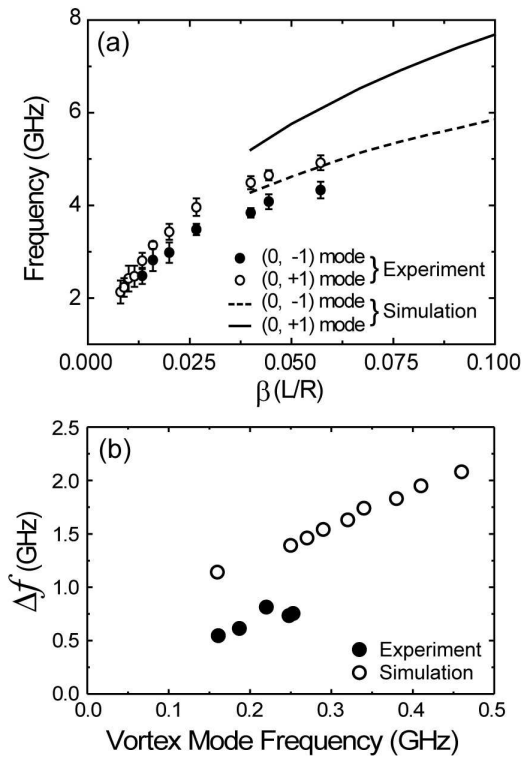}}
    \caption{(a) The measured eigenfrequencies of the (0,$\pm$1) modes are shown as a function of $\beta= L/R$.  The symbols are experimental data and the curves are the results of micromagnetic simulations. (b) The splitting $\Delta f$ of the (0,$\pm$1) modes is shown as a function of the vortex mode frequency. }
    \label{fig:fig4}
    \end{figure}
As shown by the analysis of Fig.~\ref{fig:fig3}, the symmetry of the (0,$\pm 1$) modes is determined by the polarity of
the vortex core.  The polarity determines the sense of rotation for the vortex gyrotropic motion, which breaks the
azimuthal symmetry.  Figure \ref{fig:fig4}(a) shows the eigenfrequency of (0, $\pm$1) modes as a function of the
dimensionless quantity $\beta=L/R$, where the thickness $L$ is 20 nm and the radius $R$ varies from 350 nm to 2.5
$\mu$m. The solid and dashed curves show the frequencies of the two modes determined from micromagnetic simulations. As
the radius increases, the eigenfrequencies of the (0, $\pm$1) modes and the splitting $\Delta f$  between the modes
both decrease.   For $\beta \leq 0.01$, ${\Delta}f$ becomes smaller than the frequency resolution in our experiment and
the (0,$\pm$1) modes are essentially degenerate.  Although both the absolute frequency and the size of the splitting
are larger than the experimental values, the same trend is observed in the simulations.

We have also examined the relationship between the magnitudes of the $(0,\pm1)$ splitting $\Delta f$ and the gyrotropic
mode frequency $f_G$.  This is shown in Fig.~\ref{fig:fig4}(b) for both the experiment (closed circles) and simulations
(open circles).  Over the range of our measurements, the splitting is approximately twice $f_G$, although it is not
possible to observe the gyrotropic mode in the largest disks.  The simulations considered here as well as those of Zhu
{\it et al.}\cite {Zhu:PRB2005} show a larger splitting relative to $f_G$, although they also cannot be extended to the
largest diameters, at which both $\Delta f$ and $f_G$ must approach zero.  The relationship between the vortex core mode and the azimuthal spin wave modes has also been
considered recently by Ivanov and Zaspel\cite{Ivanov:PRL2005}. They calculate a splitting $\Delta f$ which is
approximately half $f_G$, implying a curve that falls significantly below the experimental data in
Fig.~\ref{fig:fig4}(b).  Although experiment, simulations, and the theory all indicate that $\Delta f$ and $f_G$ are of
the same order of magnitude, the discrepancies among the various approaches are certainly significant.  This reflects the difficulty in treating exactly the coupling between the two different types of excitations.  Since the coupling is magnetostatic in origin, the experimental results may be susceptible to non-ideal structure at the edges of the disk.  This may also be addressed by carefully examining the role of the boundary conditions in the different theoretical approaches. 

The broken degeneracy of the azimuthal spin-wave modes clearly demonstrates how low-frequency excitations associated
with domain structure, of which a vortex is the simplest example, influence spin-wave dynamics.  An intriguing question
is how the coupling will evolve at larger wave vectors, corresponding to higher azimuthal eigenmodes.  Since the
magnetostatic interactions that provide the restoring force on the vortex core are long-range in nature, we expect that
the coupling will be smaller for higher-order modes.  At very large wave vectors, however, when the wavelength becomes
comparable to the exchange length, spin-wave modes may interact strongly with the vortex core itself.

This work was supported by the University of Minnesota MRSEC (NSF DMR-02-12032), NSF DMR-04-06029, the University of Minnesota Graduate School, and the Minnesota Supercomputing Institute.

%\bibliography{SpinDyn}

\end{document}